\documentstyle[12pt,fleqn]{article}
\begin{document}

\begin{center}

{\Large{\bf
Experimental investigations of demonstrational model
of generator\vskip3pt
using, presumably, energy of
\vskip5pt
physical vacuum.}
}
\vskip30pt
Yu.A.Baurov \& A.V.Chernikov\vskip15pt
{\it Central Scientific Research Institute of Machinery,\\
141070, Moscow reg., Kaliningrad, Pionerskaja str. 4.}
\vskip35pt
 ABSTRACT
\end{center}

In this article, a developed, manufactured and tested model
of a new type generator is presented, which allows to differentiate
and predict , with probability close to 1, coasting characteristics of the
rotor during its clockwise and counter-clockwise rotation. An explication
of the effect on a base of a new interaction, different from four known
ones, is provided.

\pagebreak

\section{Introduction.}

A model of the observed three-dementional physical space formation from
a finite set of one-dementional "vectorial" objects, byuons
\footnote{ In the works [1,2] the byuon objects were called
one-dementional discrete magnetic fluxes.}, has been proposed in [1,2].
The byuons ${\bf "[i]}$ may be presumed to be defined by the expression
$${\bf "[i] =  A_g}x(i), i=0,1,2,...,k,..$$
where ${\bf  A_g}$ is the cosmological vector-potential, a new fundamental
vectorial constant, $\vert {\bf  A_g} \vert = 1,95\times 10^{11}$ CGSE units;
$x(i)$ is a byuon quantum charge having the dimensions of length, it
has a minimum value $ \tilde x_0 = 2,8\times10^{-33}$ cm.

According to the articles [1-4], all masses of elementary particles are
proportional to the magnitude of $ {\bf A_g}$. Therefore, if the vector potential
$ {\bf A_g}$, say, of a solenoid, is in opposition to the vector $ {\bf A_g}$, one
would expect emergence of a new force in the region of reduced modules
of $ {\bf A_g}$, which force would push a particle and hence the whole of a
material out of this region. Terrestrial experiment and astrophysical
observations have shown this force to align itself predominantly with the
vector $ {\bf A_g}$, which has, in the second equatorial system, a right ascension
$\alpha = 270^\circ$ and declination $\delta = 30^\circ$ [5-9].

Based on a force predicted, a new motion principle of material body can be
realized [1,2,10,11], which consists in that a body being entered into a region
of physical vacuum with reduced $\vert {\bf  A_g}\vert$ due to the vector potential
of a magnetic source, is repelled from this region and, if connected rigidly
with a solenoid, for example, entrains it with itself.

Below are considered methods of strengthening the new interaction, constructional
approaches allowing to realize these methods, and results of experimental
investigations of a demonstrational model of the energy generator.

\section{ Methods of strengthening the new interaction.}

It is demonstrated in the articles [2,8] that the new force
$$ F\sim \lambda(\Delta A)\times\frac{\partial\lambda(\Delta A)}{\partial\Delta A}
\times\frac{\partial\Delta A}{\partial x},$$
where $\Delta A$ is the mean value of $\vert {\bf  A_g}\vert$ variation due to
the vector potential of the magnetic system in the region of a test body location,
$\lambda(\Delta A)$ is a series of $\Delta A$. It was found [2,8] that
$\lambda(\Delta A) = \sqrt{1 - v^2/c^2}$ where $c$ is the light speed in vacuum,
and $v$ is the velocity of motion of an object {\bf 4b} ("four-byuons") coming
into existence in the course of the four-contact interaction of byuons.
By the object {\bf 4b} can be meant a pair neutrino-antineutrino (an electron one).
It is precisely these objects that form the Universe [1,2].

A connection between $\lambda(\Delta A)$ and $v$ points to the fact that during
the motion if an object affecting $ {\bf A_g}$ by its own potential, in the
direction of the vector $ {\bf A_g}$ an increase in $F$ must take place.

The paper [12] reviews experimental investigations of the dependence $F = F(v_0)$
performed on a vacuum centrifugal stand when bringing magnetic rotors with
different diameters up to speed. It was shown that increase in peripheral
rotational velocity of rotors (the angular velocity kept constant)
the new force acting on the said magnetic rotors increases, too.

In the course of formation of physical space and elementary particles from
byuons [1,2], the most of their potential energy goes  into the rotational
energy of quantized objects ({\bf 4b}-ones and spins of elementary particles [2]),
as well as into that of macroscopic objects, - planets, galaxies and the entire
Universe (Birch's anisotropy [13]), but the greater part of this rotational
energy is in spins of the free objects {\bf 4b} which have Heisenberg uncertainty
interval in a coordinate being equal to $ 10^{28}$ cm [1,2] and, as a mention
above, form the space of the Universe.

Thus we may formulate the following main principle of construction of the new
force increase mechanism which consists in phase coincidence of the body motion
with the process of physical space formation from a finite set of byuons, i.e.
the body must rotate and move in the direction required. This direction is
indicated by the vector $ {\bf A_g}$.

There exist in nature some analogues of the considered principle of the new
force increasing. Firstly, this is the surfing, and secondly, the effect of
"Landau attenuation" is. If the thermal velocity of plasma electrons is
sufficiently high, an electromagnetic wave attenuates rapidly in such a plasma
transferring its energy to the electrons.

All the aforesaid can be realized, however, solely on condition that one will
be able to penetrate into the formation process of physical space, of elementary
particle charge numbers or masses. Otherwise the space observed will be isotropic
and homogeneous as it must according to the existing classical and quantum field
theory, and there will be no direction selected.

In our models of the generator, we create a "wave" in space artificially
ourselves, with the aid of the vector potential of magnetic system through
reducing $\vert  {\bf A_g}\vert$, and take energy away of the Universe's physical
vacuum by phasing motion of the body with that of the said "wave".

\section{The constructional diagram of the demonstrational model of the generator.}

In Fig. 1 a constructional diagram of the demonstrational generator model is
shown which was developed and manufactured by the Central Scientific Research
Institute of Machinery and the firm "CF". The choose of this diagram was
dictated by the above-listed methods of the new force increasing and by reducing
to a minimum inductive losses during motion of the bodies in the magnetic field.
To create a vector potential (1) field a constant magnet (2) was used, which,
 together with a magnetic circuit (3), produced a magnetic flux density $B$ in the
center of the magnetic system equal to 1.1$T$ in magnitude. In Fig.1 the vector
potential (1) of the magnetic system is directed, in the region (4) (the shaded
one), oppositely to the vector ${\bf A_g} $(5) being perpendicular to the plane of the figure,
and in the region (6) the vector (1) is aligned with ${\bf A_g}$. Hence, any body will
be ejected from the region (4) in direction of ${\bf A_g}$. Around the magnet (2) a cylindrical
drum (7) is situated, in which six bronze rollers (8) are arranged being able to
revolve the magnet (2) as the drum rotates clockwise or anticlockwise. The mass of
a bronze roller is about 1kg. The overall weight of the drum is near 15kg.
The experimental measurements of the magnetic field in the region of the drum have
given values between 0.2 and 3 mT, i.e. due to the magnetic circuit we managed to lower
significantly the magnetic field in the region of the rotating drum and, hence, to reduce
the losses due to Foucault currents to a minimum. The most contribution to dissipation
losses during the drum rotation is made by mechanical friction in the bearing (9) as well
as by aerodynamical friction. To minimize the latter, the drum (8) was built in the form
of a hollow cylinder. The construction of the generator is described in more detail in the
papers [14,15].

\section{The experimental procedure and results.}

The generator is considered as fully working when the drum (7), after being driven up to
speed with the aid of a special drive system and detached from it, comes into acceleration
under the action of the new force (mainly on the drum rollers). This can be realized only in
the event that the new force with increase in angular velocity $\omega$ of the drum rotation will be in excess
of the friction force associated with dissipation losses which grow with $\omega$, too.
To measure angular velocity of the drum rotation, an opticoelectronic system was developed, manufactured
and tested. This system is comprised of a sensing element measuring rotatory velocity of the generator
and a system of signal formation. The sensor operates over an infrared range. Its photodiode responds to
a signal reflected from the rotating surface divided into two sectors with different reflectivity of signal
coming from a light-emitting diode of the sensor. Further the signal from the photodiode arrives in the form
of rectangular pulses at a shaping circuit at which output it has the level of +15V.

With the aid of the opticoelectronic system it has been possible to follow the
variation of the angular velocity $\omega$ of the generator rotation in time $t$
(Fig.2). The following procedure was used. The rotor of the generator was
driven up to speed of about 2000 rpm after which the drive system (an electric
motor with a drive disc) was quickly drawn aside, and every seven seconds the
 opticoelectronic device gave the mean value of $\omega(t)$ over this time
interval.

The first start-up the generator was always accomplished in counter-clockwise
direction. Thereafter the rotation was reversed . The clockwise and anticlockwise
deceleration curves must be compared in pairs solely since in the course of time
the variation of nonmonitored parameters of the experimental set takes place
(modification of a lubricant, backing off of rollers from the surface of the
cylinder revolved asf.)

   In Fig.2 the pair curves of deceleration have the same designation. The continuous
curves correspond to clockwise rotation, the discontinuous ones do to anticlockwise.
The clockwise rotation correspond to the direction of the new force (in accordance
with Fig.1 if viewed from above). As is seen from Fig.2, the rotor deceleration time
during the clockwise rotation is by $(20\div 35)$\% above that for the counter-clockwise.
In 20\% of experiments from a series of 30 pairs we have observed a reverse effect.
As is seen from Fig.2, in the range from 1200 to 1500 rpm the most intensive
discrepancy between the deceleration curves corresponding to clockwise and anticlockwise
rotation takes place. It is in this range that the vibration in the rotor of the
generator was noted, and its drive rollers were much more engaged with the surface
of the cylinder revolved. Hence, the new force acting on rollers was more efficiently
transmitted to the rotor rotating. With time the rollers in experiments backed
off from the surface of the cylinder revolved under the action of centrifugal force,
and the effect began to disappear in a gradual manner (Fig.2, the curves $8, 8'$).
When changing the magnet (2) polarity (Fig.1) the curves underwent a reversal,
i.e.the rotor decelerated more slowly if rotating anticlockwise than clockwise.
Therewith the maximum discrepancy of the curves was at $\omega\approx(1300\div 1500)$ rpm.
Further they ran almost parallel to each other as in Fig.2. In $(15\div 20)$\% of 30 pairs
of experiments the reverse phenomenon was observed. Power demands of the drive
system corresponding to angular velocity $\omega\approx 2000$ rpm during clockwise and counter-
clockwise rotation of the rotor differed insignificantly from each other.

    Substitution of bronze rollers for magnetic ones having axial magnetization
(in this case are fully realized the new principle of motion and the mechanism
of increasing the new force) led to considerable difference in power consumption
of the drive system at a fixed rotational speed of the generator, or to sufficient
(up to 17%) difference in the maximum angular velocity at the fixed maximum power
consumption of the drive system (equal to $\approx300W$). Further increase in angular
velocity resulted in breakdowns of roller axes.

When used magnetic rollers more than in 30 pairs of experiments with the generator
of the type depicted in Fig.1, the deceleration time of the rotor during clockwise
rotation was in none of the cases less than during counter-clockwise.

    Thus the generator model presented may be considered as a demonstrational
one since it corroborates the existence of the new interaction and the realization
of mechanisms of its strengthening; but this model gives no way of reaching a
condition of self-acceleration of the rotor because of limitations in strength
characteristics.

\section{Conclusion.}

   On the base of the new interaction arising  when acting on physical vacuum
by magnetic systems through their vectorial potentials, a demonstrational model
of the generator using, presumably, energy of physical vacuum, was developed,
manufactured and tested. The experimental investigations of characteristics of
this model have shown that the construction developed allows to differentiate
and predict, with probability close to 1, coasting characteristics of the rotor
during its clockwise and counter-clockwise rotation. This fully confirms not only
the existence of a new force but the mechanisms of its strengthening, too.

   The generator model discussed in this article contains all basic elements of
ecologically clear generators of future using energy of physical vacuum. The
experiments in searching the most effective construction of the generator are
being continued.

Fig.1 The constructional diagram of the generator.

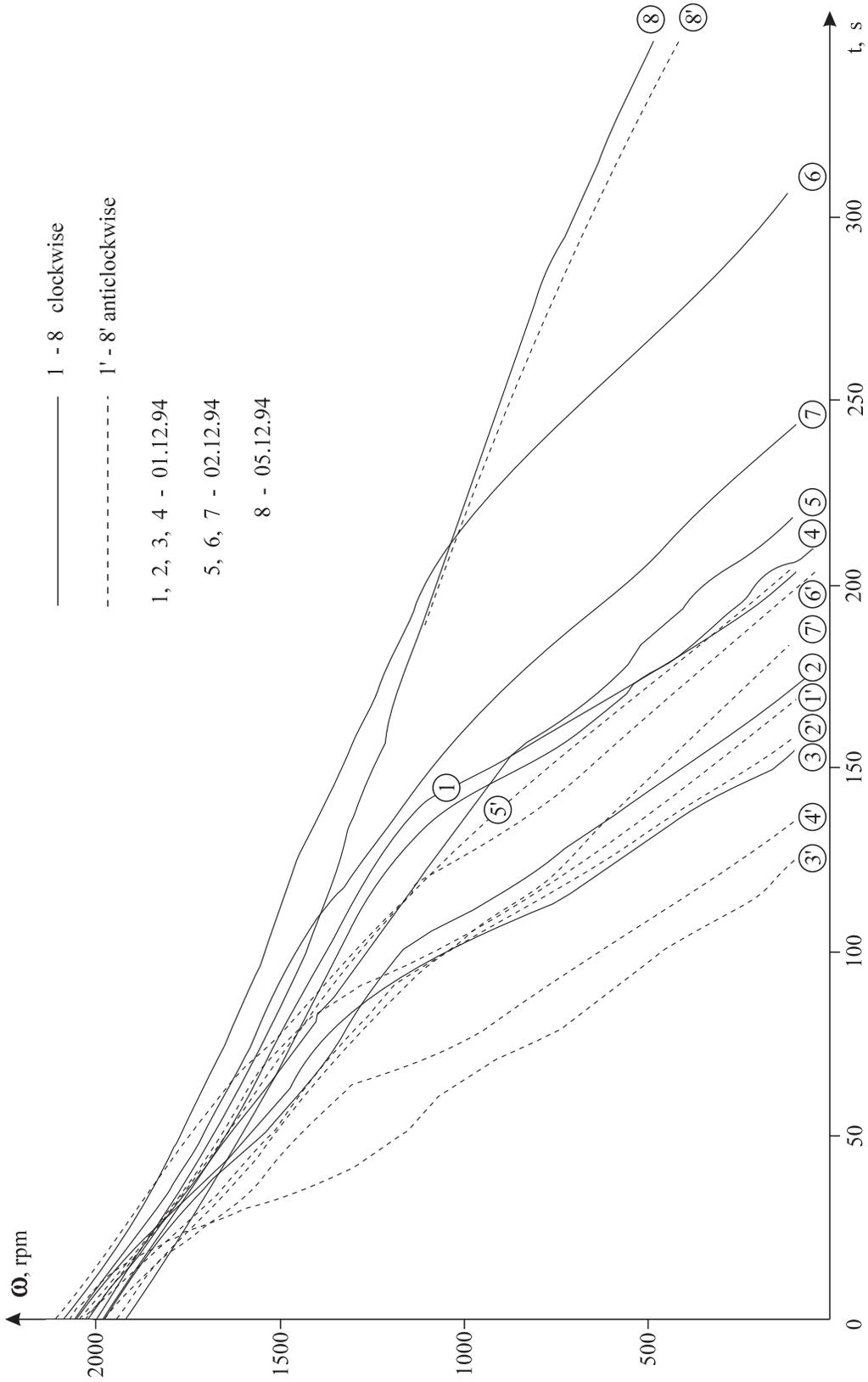
Fig.2 The dependence of the rotational speed $\omega$ on the time $t$ of rotor coasting
during clockwise and anticlockwise rotation ( $\omega$,  rpm; $t$, s;
 1-8   clockwise ,  1'-8' anticlockwise ).


\begin{thebibliography}{confr}

\bibitem{1}
Yu.A.Baurov, in collected volume "Plasma physics and some issues of general
physics", Central Scientific Research Institute of Machinery, 1990, 71-83,
84-91 (in Russian)

\bibitem{2}
 Yu.A.Baurov, Fizicheskya Mysl Rossii (Physical Idea in Russia), N1, 1994, 18-41 (in Russian)

\bibitem{3}
 Yu.N.Babajev, Yu.A.Baurov, preprint {\bf -0362} of Institute for Nuclear Researches,
Acad. of Sciences of USSR, Moscow, 1984

\bibitem{4}
 Yu.N.Babajev, Yu.A.Baurov, preprint {\bf -0368} of Institute for Nuclear Researches,
Acad. of Sciences of USSR, Moscow, 1985

\bibitem{5}
 Yu.A.Baurov, E.Yu.Klimenko, S.I.Novikov, Dokl. Akad. Nauk, v.315, N 5,
1990, 1116-1120

\bibitem{6}
 Yu.A.Baurov, E.Yu.Klimenko, S.I.Novikov, Phys.Lett., {\bf A162}, 1992, 32-34

\bibitem{7}
 Yu.A.Baurov, P.M.Ryabov, Dokl. Akad. Nauk, v.326, N1, 1992, 73-77

\bibitem{8}
 Yu.A.Baurov, Phys.Lett., {\bf A181}, 1993, 283-288

\bibitem{9}
 Yu.A.Baurov, A.A.Efimov, A.A.Shpitalnaya, Fiz. Mysl Ross., N3, 1995, 10-13 (in Russian)

\bibitem{10}
 Yu.A.Baurov, V.M.Ogarkov, Russian patent N 2023203 with priority of Nov. 11,1990

\bibitem{11}
 Yu.A.Baurov, V.M.Ogarkov, International application PCT/RU 92/00180 of
Sept. 30, 1992

\bibitem{12}
Yu.A.Baurov, V.G.Vergikovskij, Fiz. Mysl Ross., N2, 1995, 21-27 (in Russian)

\bibitem{13}
P.Birch, Nature, v.298, p.451

\bibitem{14}
A.Yu.Baurov, V.M.Ogarkov, Inventor's application N 94015479/07 (015210) of
Apr. 26, 1994

\bibitem{15}
A.Yu.Baurov, V.M.Ogarkov, International application PCT/RU 94/00135 of
June 23,1994

\end{thebibliography}
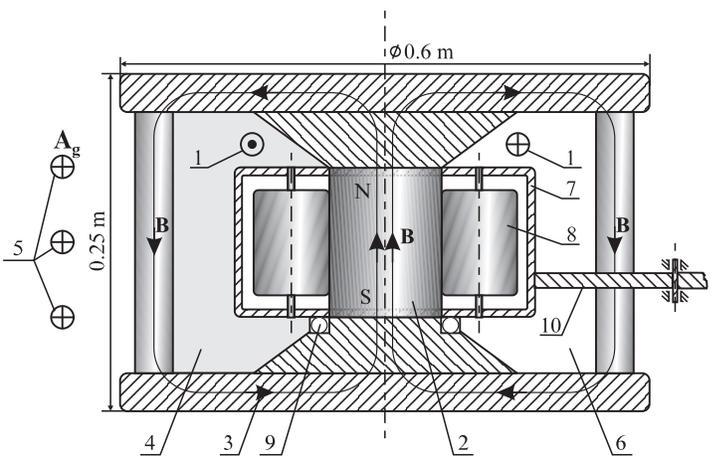
\end{document}